# Atomistic mechanism of carbon nanotube cutting catalyzed by nickel under the electron beam

*Irina V. Lebedeva ,[†,*] Thomas W. Chamberlain,[‡] Andrey M. Popov,[§] Andrey A. Knizhnik,[⊥,∥] Thilo Zoberbier,[Δ] Johannes Biskupek,[Δ] Ute Kaiser[Δ] and Andrei N. Khlobystov[‡]*

[†]Nano-Bio Spectroscopy Group and ETSF Scientific Development Centre, Departamento de Física de Materiales, Universidad del País Vasco UPV/EHU, San Sebastian E-20018, Spain, [‡]School of Chemistry, University of Nottingham, University Park, Nottingham NG7 2RD, UK, [§]Institute for Spectroscopy of Russian Academy of Sciences, Troitsk, Moscow 142190, Russia, [⊥]Kintech Lab Ltd., Kurchatov Square 1, Moscow 123182, Russia, [∥]National Research Centre "Kurchatov Institute", Moscow 123182, Russia, [Δ]Group of Electron Microscopy of Materials Science, Central Facility for Electron Microscopy, Ulm University, Albert-Einstein-Allee 11, Ulm D-89081, Germany

ABSTRACT The cutting of single-walled carbon nanotubes by an 80 keV electron beam catalyzed by nickel clusters is imaged *in situ* using aberration-corrected high-resolution transmission electron microscopy. Extensive molecular dynamics simulations within the CompuTEM approach provide insight into the mechanism of this process and demonstrate that the combination of irradiation and nickel catalyst is crucial for the cutting process to take place. The atomistic mechanism of cutting is revealed by detailed analysis of irradiation-induced reactions of bonds reorganization and atom ejection in the vicinity of the nickel cluster, showing a highly complex interplay of different chemical transformations catalysed by the metal cluster. One of the most prevalent pathways includes three consecutive stages: formation of polyyne carbon chains from carbon nanotube, dissociation of the carbon chains into single and pairs of adatoms adsorbed on the nickel cluster, and ejection of these adatoms leading to the cutting of nanotube. Significant variations in the atom ejection rate are discovered depending on the process stage and nanotube diameter. The revealed mechanism and kinetic characteristics of cutting process provide fundamental knowledge for the development of new methodologies for control and manipulation of carbon structures at the nanoscale.

INTRODUCTION

Transitions metals are widely used as catalysts for nanotube growth (see Ref. 1 for review) and graphene formation.[2] Recently the processes opposite to growth and formation, *i.e.* the etching and transformation of carbon nanostructures in the presence of transition metals, have also attracted considerable attention. These processes include metal-assisted etching of graphene[3-9] and carbon nanotubes[10-12] *via* hydrogenation[3-5] or oxidation[11,12] at high temperature and under the influence of the electron beam (e-beam) in a transmission electron microscope (TEM).[6-10] The e-beam of a TEM serves simultaneously as an imaging tool and a source of energy which activates transformations in carbon materials enabling one to study the dynamics of nanostructure growth, restructuring and etching at the atomic scale due to irradiation-induced atom ejection and bond reorganization reactions (see Ref. 13 for a review). Transition metal clusters and individual atoms play important catalytic roles in these processes, effectively lowering the activation barriers of chemical transformation driven by the e-beam. In transformations which involve the ejection of carbon atoms from the framework of the nanotubes or graphene the e-beam behaves in a similar fashion to an agent of erosion (such as oxygen) removing carbon atoms (in the form of carbon dioxide in the case of oxygen) and resulting in the formation of sidewall defects. For example, electron irradiation induced etching of graphene edges was observed as simple processes assisted by Pd,[6,7,8] Ni, Ti, Al,[7,8] or Cr[8] atoms or clusters, while iron clusters were shown to enhance the rate of graphene edge etching in the vicinity of the clusters.[9] A more complex mechanism of transformations under e-beam irradiation in single-walled carbon nanotubes containing small osmium clusters (50-60 atoms) has been observed in aberration-corrected high-resolution transmission electron microscopy (AC-HRTEM)[10]. Namely it was found that carbon atoms interacting with the osmium cluster are removed by the incident 80 keV electrons, a process that does not take place for pristine nanotubes under the same conditions, resulting in extensive defect formation followed by nanotube rupture, with the newly formed edges of the carbon nanotube observed to rearrange into closed caps after which contact between the osmium cluster and the nanotube is broken.

There is clear experimental evidence demonstrating that transition metal clusters can effectively facilitate the nanotube etching by the e-beam, but very little is known about the precise mechanism of this process. In this study we reveal the detailed mechanism of etching in carbon nanostructures catalyzed by nickel, a metal of great significance for the production, processing and practical exploitation of nanotubes and graphene. We combine experimental AC-HRTEM observations and atomistic simulations that reveal the atomic scale mechanism of carbon atom ejection facilitated by nickel under the irradiation of the e-beam.

Considerable effort has been made recently to understand the atomic-scale mechanisms of the thermally activated processes of carbon nanotube growth on iron[14-21] and nickel[19-28] catalysts and for graphene to fullerene transformations assisted by nickel clusters[29] using semi-empirical and *ab initio* atomistic simulation methods. However, examples of simulations of complex irradiation-induced processes in carbon nanostructures are much scarcer.[13,30,31] Our study combines both experimental AC-HRTEM observation of nanotube cutting and theoretical simulations, including modeling of nickel-catalyzed nanotube cutting under the action of the e-beam. We apply the recently developed CompuTEM algorithm[13,30,31] to perform molecular dynamics (MD) simulations of the atomic scale transformations of nanotubes promoted by the e-beam and catalyzed by nickel clusters. This algorithm takes into account structure relaxation between collisions with incident electrons that induce changes in the local structure and thus predicts structure evolution in real time under the experimental conditions of AC-HRTEM. The specific carbon atoms that interact with incident electrons within the simulations are classified with respect to both the number and type of chemical bonds they possess before and after electron impact. Based on this classification the relative frequencies of irradiation-induced atom ejection and bond rearrangement reactions are calculated for all types of carbon atoms. As a result a multistep atomistic mechanism which takes into account several pathways of atom ejection during nanotube cutting is proposed. The relationship between the growth and cutting mechanisms of carbon nanotubes using a nickel catalyst and the potential of this methodology to aid the elucidation of transformation processes observed for carbon nanotubes filled with clusters of different transition metals are discussed. The advantages of the studied method of nanotube cutting *via* catalyst-assisted carbon atom ejection using the e-beam of a TEM in comparison with other currently available methods of nanotube cutting[11,12,32-36] and possible applications of this new method are also discussed.

METHODS

*Materials Preparation*

Single-walled nanotubes (SWNT, arc-discharge, NanoCarbLab) were annealed at 540 °C for 20 minutes to open their termini and remove any residual amorphous carbon from the internal cavities, a 20 % weight loss was observed. Ni(hexfluoroacetylacetonate)$_2$ (10 mg) (used as supplied, Sigma Aldrich) was mixed with the SWNT (5 mg), sealed under vacuum ($10^{-5}$ mbar) in a quartz ampoule and heated at 140 °C, a temperature slightly above the vaporisation point of the metal complex, for 3 days to ensure complete penetration of the SWNT. The sample was then allowed to cool to room temperature, washed repetitively with tetrahydrofuran to remove any metal complex from the exterior of the SWNT and then filtered through a PTFE membrane (pore diameter = 0.2 µm). The nanotubes filled with metal complex was then sealed in a quartz ampoule under an argon atmosphere and heated at 600 °C, a temperature significantly above the decomposition point of the metal species (~150-200 °C), for 2 hours to decompose the metal complex into the desired pure metal nanoparticles. Alternatively the decomposition process can be achieved directly during TEM using the e-beam as the energy source. Metal particles formed by thermal and e-beam decomposition of the metal complex are virtually indistinguishable.

*TEM conditions*

AC-HRTEM imaging was carried out using an image-side $C_s$-corrected FEI Titan 80-300 transmission electron microscope operated at 80 kV acceleration voltage with a modified filament extraction voltage[37] for information limit enhancement. Images were recorded on a slow-scan CCD-camera type Gatan Ultrascsan XP 1000 using binning 2 (1024 by 1024 pixel image size) with exposure times between 0.2-1.0 s. For all in-situ irradiation experiments the microscope provided a highly controlled source of local and directed electron irradiation on a selected area of the sample. Experimentally applied electron-fluxes ranged from $2 \cdot 10^6$ to $9 \cdot 10^6$ e$^-$/nm$^2$/s, and the total applied dose was kept the same, reaching approximately $10^{10}$ e$^-$/nm$^2$ at the end of each experiment. TEM specimens were heated in air at 150 ºC for 7 min shortly before insertion into the TEM column. All imaging experiments were carried out at room temperature.

*Reactive empirical MD simulations*

Effective modelling of the processes induced by electron irradiation were achieved using the CompuTEM algorithm[13,30,31] in which only interactions between incident electrons and atoms which lead to changes in the atomic structure (i.e. irradiation-induced events) are taken into account. The structure is annealed between each induced event at elevated temperatures to take into account reorganisation of the structure between events. Carbon nanotubes are metallic or narrow-gap semiconductors, so that their ionisation and excitation cross sections are expected to be orders of magnitude lower than the cross section for elastic processes, which is analogous to other metallic or semiconducting materials.[38,39] Furthermore, because only a small segment of a nanotube is exposed to the e-beam during TEM imaging, any ionisation, excitation or heat effectively dissipate due to the excellent electric and heat conductance of carbon nanotubes. Therefore, processes triggered by kinetic energy transfer from incident electrons to atoms (knock-on) dominate the transformations observed in TEM, while ionisation, excitation and heating due to the e-beam remain insignificant. For this reason, consideration of only elastic collisions between incident electrons and atoms is adequate for simulations of carbon nanotube transformations in TEM.

According to the CompuTEM algorithm,[13,30,31] irradiation-induced events are described as follows: 1) the nanostructure is equilibrated at a temperature corresponding to the experimental conditions in AC-HRTEM, 2) the type of each atom of the nanostructure is determined based on the number and strength of its chemical bonds, 3) the possible minimal energy that can be transferred from an incident electron is assigned to each atom in accordance with the atom type determined in step 2, 4) a single electron-atom interaction event is introduced by giving momentum distributed according to the standard theory of elastic electron scattering between a relativistic electron and the nucleus[40,41] of a random atom that is chosen based on the total probabilities of electron collisions with different atoms determined by the minimum transferred energies assigned in step 3, 5) MD are performed at a temperature corresponding to the experimental conditions with a duration sufficient for bond reorganisation to occur, 6) the surroundings of the impacted atom are analysed again and if no change in the atom type or in the list of the nearest neighbours is detected as compared to step 2 within this time period (the impact is unsuccessful), the simulation cycle is repeated. However, if the system topology has changed (the impact is successful), an additional MD run of a duration of $t_{rel}$ at elevated temperature, $T_{rel}$, is performed to describe relaxation of the structure between successive electron impacts.

The assignment of atom types in step 2 of the algorithm is performed based on the following information: (1) the number of carbon neighbours the atom has and the coordination numbers of the neighbouring atoms, (2) the presence of the nearest-neighbour nickel atoms and (3) the existence of non-hexagonal rings in the carbon network of the nanotube to which the atom belongs. To simplify the choice of minimal transferred energies $E_{min}$, all atom types were divided into three groups with different values of $E_{min}$. Based on our previous studies,[13,31] the following minimal transferred energies were selected: (1) $E_{min}^{(1)} = 5$ eV for carbon adatoms and ad-dimers adsorbed or dissolved in the nickel cluster and for one-coordinate carbon atoms, (2) $E_{min}^{(2)} = 10$ eV for two-coordinate atoms and three-coordinate atoms in non-hexagonal rings or within two bonds from one-coordinate, two-coordinate and three-coordinate atoms in non-hexagonal rings, and (3) $E_{min}^{(3)} = 17$ eV for three-coordinate atoms in the perfect hexagonal part of the carbon network (two bonds away from any atoms of the types listed above). As most of the irradiation-induced events take place for atoms from the second group, two values for the corresponding minimal transferred energy $E_{min}^{(2)} = 10$ eV and 13 eV are considered in the present paper to study the sensitivity of the results to this parameter. Electron impacts that occur with the nickel cluster are disregarded due to the large mass of the nickel atoms.

The use of accurate interatomic potentials is indispensable for precise modeling of structural transformation induced by electrons with a kinetic energy of 80 keV. Therefore, we described all interatomic interactions by the new potential for nickel-carbon systems which was recently elaborated on the basis of the first-generation bond-order Brenner potential.[29] This potential reproduces both experimental and first-principles data on the physical properties of pure nickel as well as the relative energies of carbon species on nickel surfaces and in bulk nickel metal.

An in-house MD-kMC[42] (Molecular Dynamics – kinetic Monte Carlo) code was used. The integration time step was 0.6 fs. The temperature is maintained by the Berendsen thermostat,[43] with relaxation times of 0.1 ps, 3 ps and 0.3 ps for the MD runs in steps 1, 5 and 6 of the described algorithm, respectively. To identify non-hexagonal rings the topology of the carbon bond network of the nanotube is analyzed on the basis of the "shortest-path" algorithm.[44] Two carbon atoms are considered as bonded if the distance between them does not exceed 1.8 Å, while for bonded carbon and nickel atoms, the maximum bond length is 2.2 Å. As the finite size of the simulation box can lead to artificial reattachments of emitted atoms and dimers (they can cross the simulation box and stick back), atoms and dimers that detach from the system and do not stick back within 10 ps were removed.

RESULTS AND DISCUSSION

*Experiment*

Ni metal was encapsulated in carbon nanotubes in the form of nickel hexafluoroacetylacetonate, $Ni(C_5HF_6O_2)_2$, which can be easily broken down into pure metal and ligand.[45] While the Ni atoms aggregate into clusters of 50-100 atoms forming intimate contact with the nanotube inner (concave) surface, the ligand is broken into small fragments and leaves the nanotubes. The identity of the metallic clusters formed inside the nanotubes was confirmed by energy dispersive X-ray (EDX) spectroscopy using a focused 100 keV electron beam on a small bundle of 5-10 filled SWNTs (ESI file). The evolution of metal clusters and their interactions with carbon sample was analysed using AC-HRTEM at 80 keV using an electron flux between $10^6$-$10^7$ e$^-$/nm$^2$/s and a cumulative dose for each image series of *c.a.* $10^{10}$ e$^-$/nm$^2$. This energy of the e-beam is insufficient for direct ejection of carbon atoms from a defect-free nanotube,[46] which allows the evaluation of the effect of the nickel clusters on transformations of nanotubes under the e-beam. Time series of images recorded for individual nickel clusters consistently show strong interaction between the nickel clusters and the nanotube, manifested in bonding of the clusters to the interior of SWNT observed in AC-HRTEM images, which results in extensive transformation of the nanotube structure and eventual nanotube cutting, Figure 1.

For some applications in nanodevices it is necessary to control the cutting of a single-walled nanotube positioned between electrodes to create a small tunable gap between the ends of cut nanotube.[47-49] However most of the present techniques for cutting of individual nanotubes are not able to form reproducibly gaps smaller than 10 nm[47-52] in the nanotube structure and can lead to contamination of the interior of the nanotubes with carbon debris[53] that can affect the electronic properties of nanotubes. In contrast, the nickel-cluster-catalysed cutting of nanotubes by the electron beam in HRTEM investigated here is a method which can in principle enable cutting a nanotube after nanodevice assembly with atomic precision, reproducibly creating the smallest gaps in the SWNT structure dictated by the size of the nickel cluster of about 1 nm.

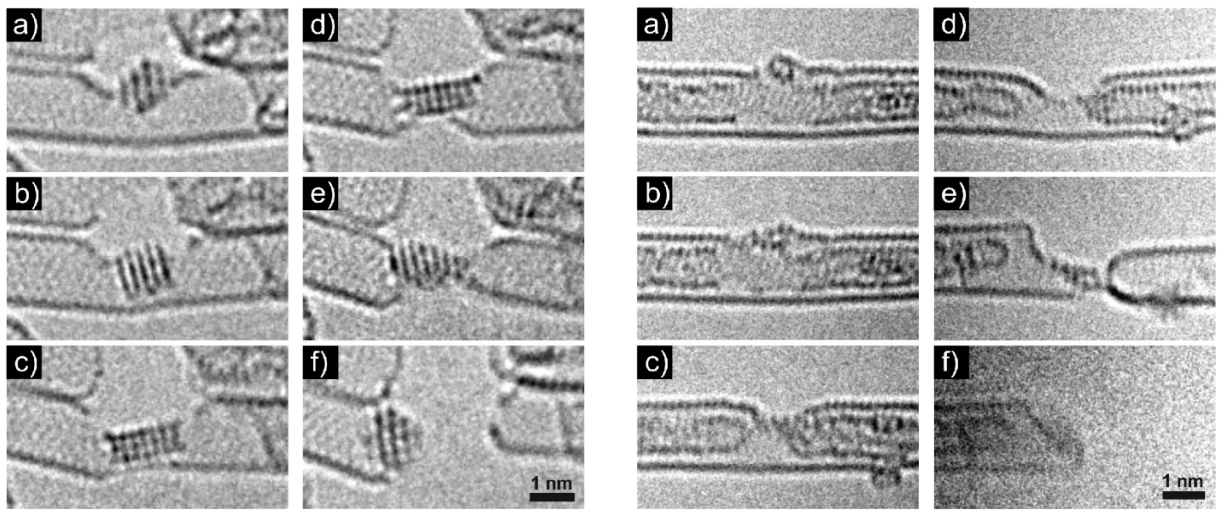

**Figure 1.** Time series of consecutive experimental AC-HRTEM images (a-f) showing the key stages of carbon nanotube cutting catalysed by a nickel cluster. Experimental details: (Left) total time 400s + 40s offset (for searching, focusing, stigmation), dose $1.44 \cdot 10^9$ e$^-$/nm$^2$ (time series only) and $1.6 \cdot 10^9$ e$^-$/nm$^2$ (including the offset); (Right) total time 113s + 40s, dose $0.18 \cdot 10^9$ e$^-$/nm$^2$ (time series only) and $0.25 \cdot 10^9$ e$^-$/nm$^2$ (including the offset).

*Reactive empirical MD simulations*

**Carbon nanotube cutting by nickel clusters under electron irradiation.** To investigate the detailed mechanism and kinetics of the nickel-assisted cutting of carbon nanotubes we performed reactive empirical MD simulations of this process based on the recently developed CompuTEM algorithm.[13,30,31] A (5,5) carbon nanotube 43 nm in length is considered (Figure 2a) in which the initial structure of the nanotube is geometrically optimized. The two-coordinate edge carbon atoms of the nanotube (in the present paper, only the nearest-neighbour carbon atoms are included in the coordination number) are fixed to prevent displacement of the nanotube. Six neighbouring carbon atoms are removed from the central part of the nanotube to form a hole on which the Ni$_{13}$ cluster is adsorbed to initiate nanotube cutting. This is consistent with most experimentally observed processes in which the nickel clusters are often observed to be already adsorbed on pre-existing defects in the nanotube sidewall at the start of the cutting transformation (vacancy type defects are very common in SWNTs as defect-free nanotubes are virtually non-existent). The kinetic energy of all incident electrons and the electron beam flux are set at 80 keV and $4.1 \times 10^6$ electrons·s$^{-1}$·nm$^{-2}$ respectively, to match the experimental conditions used for AC-HRTEM imaging.

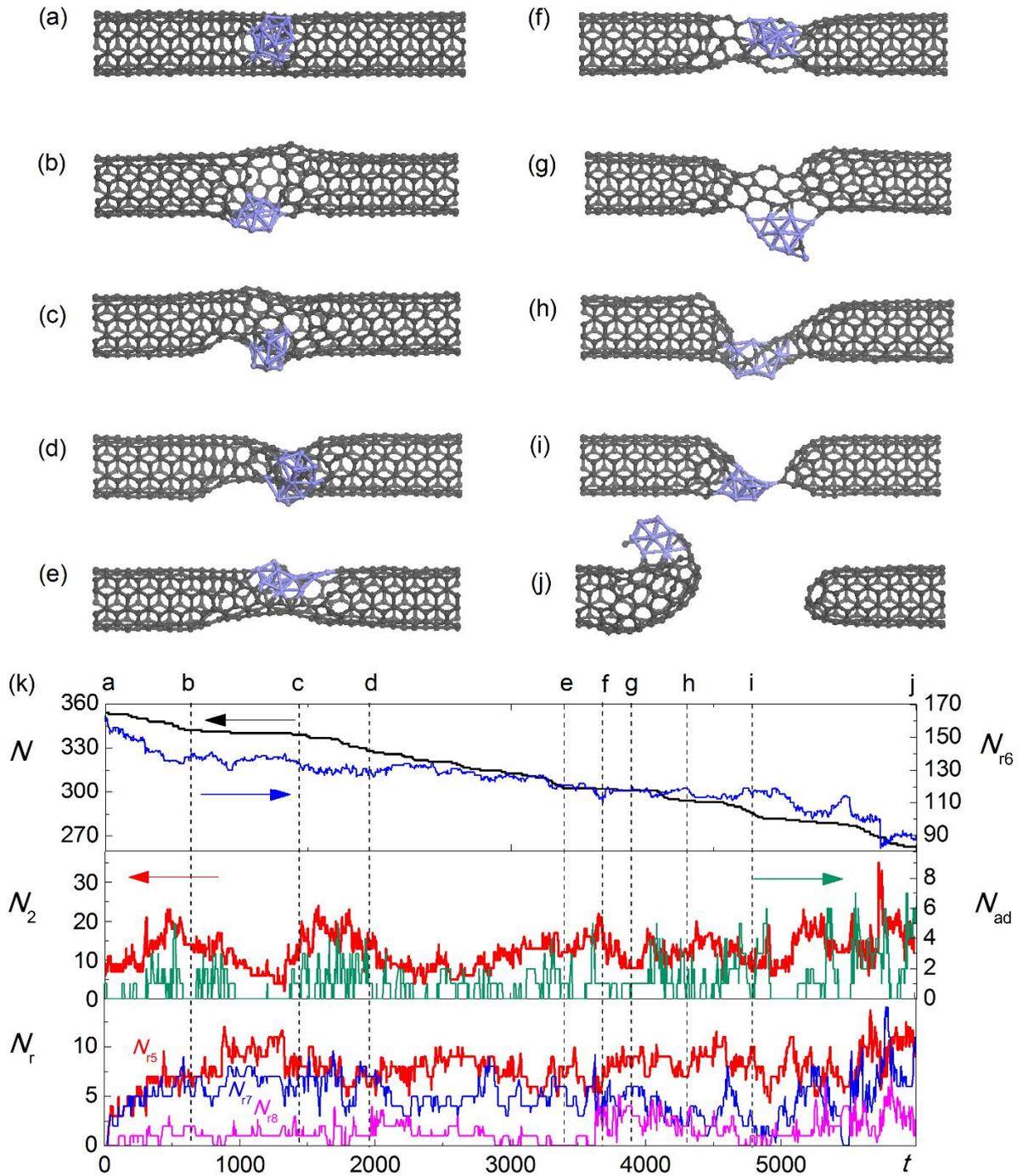

**Figure 2.** (a-j) Simulated evolution of the structure of a carbon nanotube with an adsorbed nickel cluster under irradiation by electrons with kinetic energy of 80 keV and a flux of $4.1 \cdot 10^6$ electrons/(s·nm$^2$): (a) 0 s, (b) 637 s, (c) 1441 s, (d) 1968 s, (e) 3397 s, (f) 3688 s, (g) 3900 s, (h) 4305 s, (i) 4797 s, (j) 6070 s. The direction of the electron beam is out of the page. (k) Calculated total number of carbon atoms, $N$, in the considered structure (black line, left axis, upper panel), number of hexagons, $N_{r6}$ (blue line, right axis, upper panel), number of two-coordinate and one-coordinate atoms, $N_2$ (thick red line, left axis, middle panel), number of carbon atoms and dimers adsorbed or dissolved in the nickel cluster, $N_{ad}$ (thin green line, right axis, middle panel) and the numbers of pentagons, heptagons and octagons, $N_{r5}$, $N_{r7}$ and $N_{r8}$ respectively (red, blue and magenta lines, respectively, lower panel) as functions of time, $t$ (in s). The moments of time corresponding to structures (a–j) are shown using vertical dashed lines. The minimal transferred energy is $E_{min}^{(2)} = 13$ eV and the duration of high-temperature relaxation between electron collisions is $t_{rel} = 30$ ps (series A).

In addition to revealing the atomistic mechanism of nanotube cutting, in this study we also test the key parameters of simulations which involve carbon systems under electron irradiation with the help of the CompuTEM algorithm. In our previous studies which modelled the irradiation-induced transformation of graphene to fullerene for an all-carbon system[13] and in the presence of a nickel cluster,[31] we chose the following simulation parameters that adequately described this process: (1) a minimal energy, $E_{min}^{(2)}$ = 10 eV, which is introduced to avoid description of low-energy electron impacts that do not induce structural transformations, was transferred to two-coordinate and three-coordinate carbon atoms in non-hexagonal rings of the carbon bond network or within two bonds from one-coordinate, two-coordinate and three-coordinate carbon atoms in non-hexagonal rings (as shown below most irradiation-induced events involve these types of atoms, see section "Methods" for the values of minimal transferred energy for other types of atoms); (2) the temperature of the MD stage describing the structure relaxation between successive irradiation-induced events, $T_{rel} = 1800 - 2500$ K; and (3) the duration of the relaxation stage, $t_{rel} = 100$ ps. However, simulations with such parameters become too consuming for processes which involve in the order of 300 irradiation-induced events for a system consisting of almost 400 atoms studied here. Therefore, we considered the possibility of accelerating the simulations by decreasing the duration of the high-temperature relaxation stage, $t_{rel}$, and increasing the minimal transferred energy $E_{min}^{(2)}$. Higher values for the minimal transferred energy help to increase the portion of successful impacts which lead to local structure changes. Therefore, to study the sensitivity of the results to changing these simulation parameters we performed two series of 10 simulations with different parameters. In series A the minimal transferred energy $E_{min}^{(2)}$ = 13 eV, and the structure relaxation step duration, $t_{rel} = 30$ ps at a temperature, $T_{rel} = 2000$ K. In series B the minimal transferred energy, $E_{min}^{(2)} = 10$ eV, and the duration, $t_{rel} = 10$ ps at a temperature, $T_{rel} = 2000$ K.

These series of simulations both correlate well with the experimental observations that upon 80 keV electron irradiation the nickel cluster cuts the carbon nanotube into two parts closed by caps separated by a distance of approximately 10 Å (Figure 2j) and remains adsorbed on one of the SWNT caps at the end of the process. The loss of connectivity between the two nanotube caps formed as a result of cutting of the initial nanotube was observed within 8700 ± 900 s in 9 simulations of series A and 5300 ± 500 s in 9 simulations of series B. The average number of ejected carbon atoms at the moment of separation of the nanotube into two non-interacting parts is 93 ± 5 and 83 ± 2 for the completed simulations of series A and B, respectively. The average time between carbon atom ejection events is calculated to be 99 ± 7 s and 74 ± 5 s in series A and B (**Table 1**), respectively, in which is comparable to the previous simulations of irradiation-induced graphene-fullerene transformations assisted by nickel clusters.[31] Though there is some quantitative difference in the kinetics of the process in series A and B, no qualitative differences are observed in the evolution of the structures in these two series. Therefore, the parameters chosen for the simulations in this study can be considered adequate to investigate the process of carbon nanotube cutting (this is discussed in more detail in the Supplementary Information).

**Table 1.** Calculated average time between different irradiation-induced events, their relative frequencies for impacted carbon atoms of different types and the number of atoms of different type averaged over total time for the simulation of nickel catalyzed electron beam assisted nanotube cutting with a MD relaxation step temperature of $T_{rel} = 2000$ K.

|  | Irradiation-induced events | | | | Number of atoms[a] | |
|---|---|---|---|---|---|---|
|  | All | | Ejection of atoms | | | |
| Simulation series | A | B | A | B | A | B |
| Minimal transferred energy $E_{min}$ (eV) | 13 | 10 | 13 | 10 | 13 | 10 |
| Relaxation time $t_{rel}$ (ps) | 30 | 10 | 30 | 10 | 30 | 10 |
| Total number of irradiation-induced events | 33491 | 50434 | 849 | 797 |  |  |
| Average time between ejection events $\tau$ (s) | 2.550 ± 0.016 | 1.116 ± 0.008 | 99 ± 7 | 74 ± 5 |  |  |
| Atom types | Not bonded to the cluster | | | | | |
| One-coordinate atoms | 0.0008 | 0.0006 | 0.0094 | 0.0075 | 0.02 | 0.02 |
| Two-coordinate atoms except atoms in chains | 0.0222 | 0.0196 | 0.0306 | 0.0163 | 3.74 | 4.10 |
| Two-coordinate atoms in chains[b] | 0.0123 | 0.0115 | 0.0035 | 0.0037 | 1.81 | 2.11 |
| Three-coordinate atoms in non-hexagonal rings | 0.6228 | 0.6180 | 0.0200 | 0.0113 | 84.69 | 84.64 |
| Three-coordinate carbon atoms in hexagons | 0.1660 | 0.1564 | 0.0012 | 0.0025 | 171.89 | 169.64 |
| Total for atoms not bonded to the cluster | 0.8241 | 0.8062 | 0.0648 | 0.0414 | 262.14 | 260.51 |
| | Bonded to the cluster | | | | | |
| One-coordinate atoms | 0.0295 | 0.0225 | 0.1284 | 0.1468 | 0.55 | 0.86 |
| Two-coordinate atoms except atoms in chains | 0.0504 | 0.0468 | 0.0636 | 0.0464 | 4.14 | 4.03 |
| Two-coordinate atoms in chains[b] | 0.0511 | 0.0652 | 0.0766 | 0.1004 | 4.09 | 5.26 |
| Three-coordinate atoms in non-hexagonal rings | 0.0082 | 0.0097 | 0 | 0 | 0.49 | 0.55 |
| Three-coordinate carbon atoms in hexagons | 0.0026 | 0.0037 | 0.0012 | 0 | 0.15 | 0.18 |
| Adatoms | 0.0720 | 0.0427 | 0.6525 | 0.6487 | 1.26 | 1.59 |
| Ad-dimers | 0.0039 | 0.0034 | 0.0118 | 0.0163 | 0.03 | 0.06 |
| Total for atoms bonded to the cluster | 0.2177 | 0.1938 | 0.9352 | 0.9586 | 10.71 | 12.54 |

[a] Fixed atoms at the nanotube edges are not counted.
[b] Two-coordinate carbon atoms which have at least one bond with two-coordinate or one-coordinate carbon atoms.

In both series of simulations, nanotube cutting proceeds through several stages. Firstly, the loss of carbon atoms around the nickel cluster results in the growth of the hole in the nanotube sidewall (Figure 2b), followed by rearrangement of the carbon atoms around the edge of the defect minimizing the number of dangling bonds, *i.e.* sewing up of the hole, which gives rise to a local decrease of the nanotube diameter (Figure 2c). The repetition of growing/sewing cycles for the hole leads to its diffusion (and diffusion of the nickel cluster attached to the hole edges) on the nanotube sidewall. As the ejection of carbon atoms can take place not only along the nanotube circumference but also in the direction parallel to the nanotube axis, the hole with the nickel cluster attached to its edges can slowly diffuse along the nanotube wall leading to an increase in the length of the narrow section of the nanotube (Figure 2d). Eventually, through further loss of carbon atoms, the nanotube becomes narrower and narrower until locally its diameter is so small that this region of the nanotube unfolds into a graphene nanoribbon (Figure 2e). This happens when the elastic energy of this region becomes comparable to the energy of the unterminated edges of the graphene nanoribbon. Fluctuations between structures in which the curved graphene nanoribbon has all carbon edges attached to the nickel cluster and the flat nanoribbon with partially free carbon edges are observed. Further loss of carbon atoms around the nickel cluster leads to structures with only carbon chains and discrete rings left bridging between the two capped nanotubes (Figure 2f). These intermediate stages are not strictly sequential and reconstruction of the graphene nanoribbon is also observed (Figure 2g), followed by decomposition into atomic chains (Figure 2h). At some stages, only the nickel cluster keeps the nanotubes in contact (Figure 2i). However, if the separation between the nanotube caps is not sufficient ejection of carbon atoms from the nanotubes continues and can be accompanied by a reappearance of chains of carbon atoms connecting the nanotubes. Finally the cluster detaches from one of the nanotube caps and ejection of carbon atoms from this nanotube cap then stops (Figure 2j).

During the simulations, the carbon bond network was analyzed in depth. The total number of atoms, $N$, the number of two-coordinate and one-coordinate carbon atoms, $N_2$, the number of carbon atoms in non-hexagonal rings, $N_d$, the number of carbon atoms dissolved in or adsorbed on the nickel cluster, $N_{ad}$ (with no bonds to other carbon atoms), and the number of carbon rings of different size, $N_{r5}$, $N_{r6}$, $N_{r7}$ and $N_{r8}$, were monitored (Figure 2k). The gradual loss of carbon atoms can be seen by a decrease in the total number of carbon atoms in the system, $N$, and hexagons, $N_{r6}$, in the carbon network (Figure 2k). Fluctuations in the perimeter of the hole in the carbon nanotube manifest themselves in changes in the total number of two-coordinate and one-coordinate carbon atoms, $N_2$. A considerable number of carbon atoms detach from the nanotube wall and dissolve in the nickel cluster. Figure 2k shows that the number of carbon atoms adsorbed and dissolved in the metal cluster, $N_{ad}$, also exhibits significant fluctuations. It is also observed that the number of non-hexagonal rings ($N_{r5}$, $N_{r7}$ and $N_{r8}$) increases rapidly at the beginning of the simulations and stabilizes within 500 s. The formation of near perfect nanotube end caps can be observed shortly before the final cutting as only 2-3 heptagons and octagons are present in the structure of both ends of the nanotube (Figure 2i, k). Overall, continuous ejection of carbon atoms under irradiation with the electron beam results in the constant introduction of topological defects into the nanotubes (Figure 2j, k).

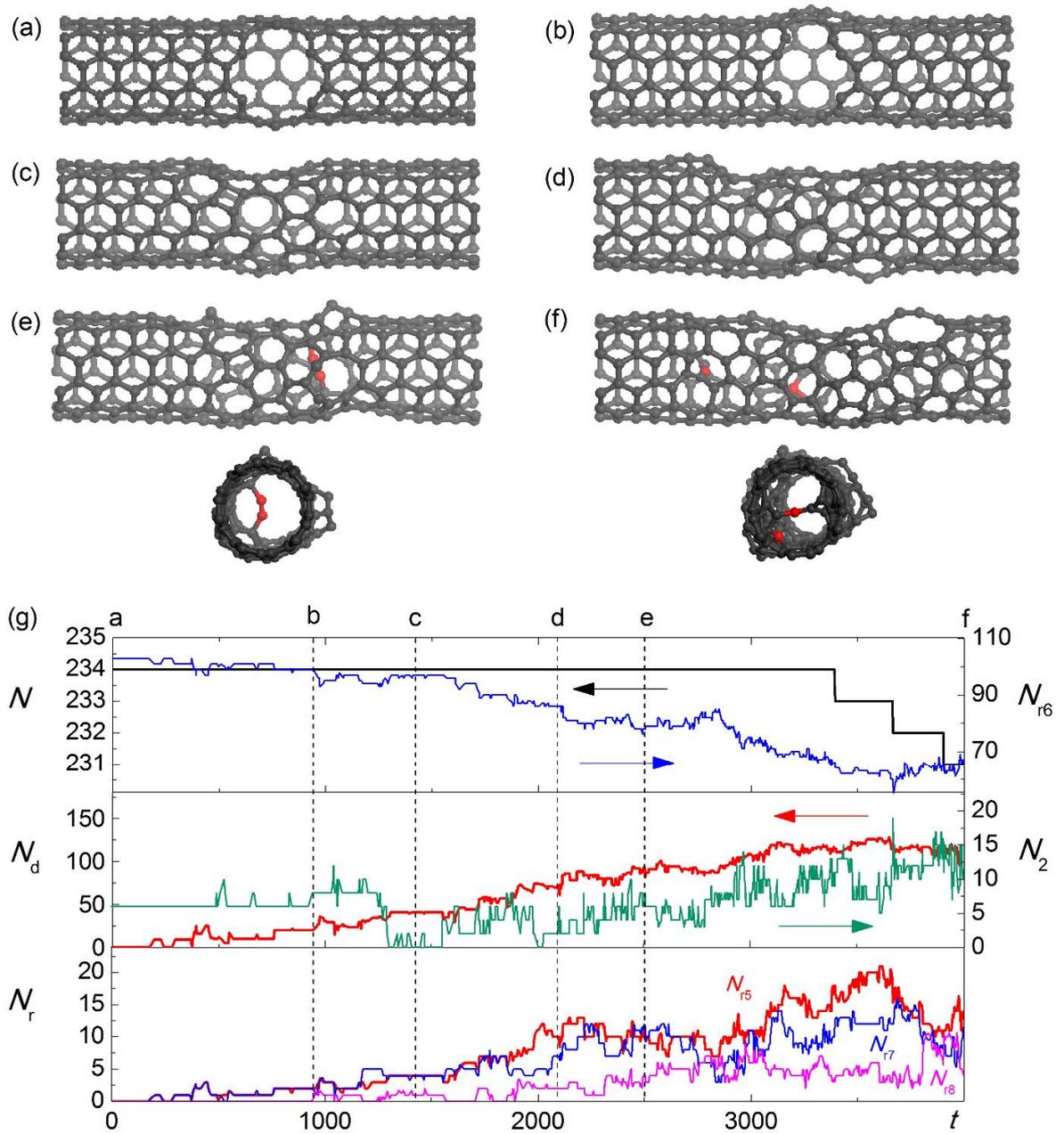

**Figure 3.** (a-f) Simulated evolution of the structure of a carbon nanotube with a hole of 6 atoms under irradiation by electrons with a kinetic energy of 80 keV and a flux of $4.1 \cdot 10^6$ electrons/(s·nm$^2$): (a) 0 s, (b) 944 s, (c) 1419 s, (d) 2092 s, (e) 2503 s, (f) 4000 s. (a-d) Side view, the direction of the electron beam is out of the page. (e,f) View from the side and along the nanotube axis. (g) Calculated total number of carbon atoms, $N$, in the considered structure (black line, left axis, upper panel), number of hexagons, $N_{r6}$ (blue line, right axis, upper panel), number of atoms in non-hexagonal rings, $N_d$ (thick red line, left axis, middle panel), number of two-coordinate and one-coordinate atoms, $N_2$ (thin green line, right axis, middle panel), and the numbers of pentagons, heptagons and octagons, $N_{r5}$, $N_{r7}$ and $N_{r8}$ respectively (red, blue and magenta lines, respectively, lower panel) as functions of time, $t$ (in s). The moments of time corresponding to structures (a−f) are shown using vertical dashed lines. The minimal transferred energy is $E_{min}^{(2)} = 13$ eV and the duration of high-temperature relaxation between electron collisions is $t_{rel} = 30$ ps (the same as in series A).

Detailed analysis of the structure of the nanotube end caps in the simulations of series A, just after a cutting step, reveals that at the moment of separation of the two segments the nanotube, contains on average 16.0 ± 1.1 pentagons, 11.4 ± 1.3 heptagons and 2.4 ± 0.8 octagons. The nanotubes also have a considerable number of two-coordinate and one-coordinate carbon atoms, distributed as 8.0 ± 2.0 two-coordinate carbon atoms in carbon chains, 5.1 ± 1.1 two-coordinate carbon atoms not in chains, and 0.7 ± 0.3 one-coordinate carbon atoms. In series B, the nanotubes at the moment of separation contain 14.2 ± 1.3 pentagons, 10.6 ± 0.9 heptagons, 2.0 ± 0.4 octagons, 6.4 ± 2.4 two-coordinate carbon atoms in chains, 6.1 ± 1.4 two-coordinate carbon atoms not in chains and 0.7 ± 0.2 one-coordinate carbon atoms. These numbers illustrate that after the cutting step both nanotube caps still contains holes and topological defects, and the nickel cluster remains adsorbed on a hole in one of the caps. It is expected that further electron irradiation of the two segments of nanotube should result in their reconstruction and decrease of defects.

**E-beam irradiation of pristine nanotubes and thermal treatment of carbon nanotubes with nickel clusters.** To consider the effects of the metal cluster and electron irradiation separately we carried out supplementary simulations in the absence of each of these factors individually. Two series of simulations in the absence of the nickel cluster were performed with the same parameters as in series A and B. However, the structure was found to evolve in a very similar fashion in these two series. The ejection rate observed in simulations for the pristine nanotube without the nickel cluster adsorbed (Figure 3a) is an order of magnitude smaller than the ejection rate in the presence of the nickel cluster. Furthermore, the structural reconstruction of the nanotube induced by the e-beam in this case proceeds *via* alternative pathways (Figure 3). Except for short-living metastable states corresponding to temporary elimination of the hole in the nanotube sidewall (Figure 3c), the number of non-hexagonal rings in these simulations is constantly growing. The number of one and two-coordinate carbon atoms is also increasing steadily and exhibits significant fluctuations (Figure 3g), resulting in opening and closing of holes in the nanotube sidewall (see, for example, a hole on the reverse side of the nanotube in Figure 3f). In addition to non-hexagonal rings, bridges between opposite sides of the nanotube are formed by individual or chains of two-coordinate carbon atoms (Figure 3e and f).

Simulations of the same initial structure with a nickel cluster adsorbed on the hole at temperature 2000 K without electron irradiation did not reveal any structural rearrangements within tens of nanoseconds. In particular, no atom ejection was observed. Therefore, a combination of both electron irradiation and a metal cluster, is required to initiate nanotube cutting using electrons of 80 keV energy (see Supplementary Information for more details).

**Atomistic mechanism of nanotube cutting.** Previously theoretical consideration of e-beam initiated carbon nanotube cutting catalysed by osmium clusters focused simply on the formation of the initial defect in the pristine structure of the nanotube.[10] To deduce the mechanism of atom ejection at the subsequent stages of the cutting process we considered the statistics of events induced by irradiation (dissociation and rearrangement of chemical bonds, ejection of carbon atoms etc.) for different types of carbon atoms depending on their neighbours within the carbon network: one-coordinate atoms, two-coordinate atoms within chains, two-coordinate atoms excluding those in chains, three-coordinate atoms in non-hexagonal rings and three-coordinate atoms in hexagons. Additionally, whether the carbon atoms are bonded to the nickel cluster or not was also analysed. Where the carbon atoms were bonded to the nickel cluster a further distinction was made between carbon adatoms and ad-dimers of atoms ($C_2$), *i.e.* atoms and dimers not bonded to the nanotube network (N.B. any single carbon atoms and dimers not bonded to the cluster are considered to have been knocked-out and are thus removed from the system as described in the "Methods" section).

Table 1 shows that 80% of events are induced in carbon atoms that are not bonded to the nickel cluster (see section "Methods" for the definition of bonded atoms). However, these atoms are still located in the vicinity of the cluster and non-perfect regions of the nanotube (such as a hole in the nanotube wall). This is in agreement with the previous observation that no reactions are induced in perfect nanotubes by irradiation of electrons with an energy of 80 keV. The majority of reactions, 60%, involve three-coordinate carbon atoms in non-hexagonal rings. These reactions are responsible for reconstruction at the ends of the two nanotube sections formed upon cutting and are discussed below. However, only 4-7% of atom ejection events, which represents only a small fraction of all irradiation-induced events (2.5% and 1.6% for series A and B, respectively), occur with carbon atoms that are not bonded to the nickel cluster, with almost all atom ejection events taking place with carbon atoms bonded to the metal cluster (**Table 1**), thus proving that the metal cluster plays a crucial role in nanotube cutting.

Detailed analysis of several different pathways for nickel cluster assisted carbon atom ejection revealed the ejection of adatoms, two-coordinate and one-coordinate carbon atoms bonded to the cluster as the major mechanisms. The main pathway corresponds to ejection of single carbon atoms adsorbed or dissolved in the nickel cluster (about 65% of emitted atoms). Some formation of ad-dimers and their subsequent expulsion from the cluster are also observed, though the contribution of this pathway to the overall process is very small (about 1-2%). It should be mentioned that in the simulations at 2000 K without electron irradiation no carbon atoms get detached from the nanotube sidewall and dissolve in the cluster. Therefore, formation of adatoms and ad-dimers must also be a result of irradiation-induced events. In particular, there appears to be a link between the number of carbon atoms dissolved in or adsorbed on the cluster and the combined number of carbon atoms in chains and one-coordinate carbon atoms bonded to the nickel cluster. The ratio of these two numbers averaged over time is close to 0.26 in all the simulations both for series A and B, with the relative root-mean-square deviations for the simulations in these series only 0.09 and 0.06, respectively. Therefore, chains of carbon atoms bonded to the nickel cluster can be considered as a primary source of carbon adatoms on the cluster.

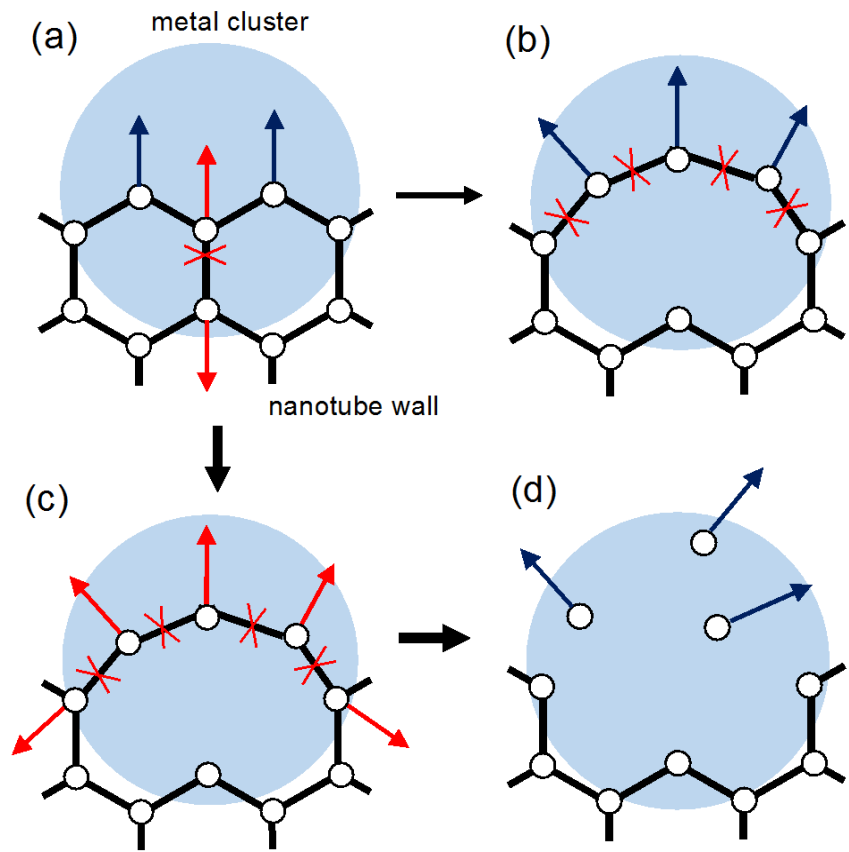

**Figure 4.** Scheme showing the proposed mechanism of carbon atom ejection: (a,c,d) the main pathway *via* transfer of carbon atoms to the metal cluster followed by ejection of carbon adatoms (d), (a,b) additional pathway - ejection of two-coordinate atoms from carbon chains (b) and at the edge of a hole (a). The metal cluster is indicated by the blue circle, carbon atoms and bonds between them are shown by white circles and black lines, respectively. Bond breaking is shown by red crosses. Atoms that experience electron impacts and directions of momentum transfer

from electrons that facilitate this bond breaking are indicated by red arrows. The ejection of atoms is indicated by dark blue arrows. Ejection of one-coordinate carbon atoms formed at the intermediate stages which are not shown in the figure (i.e. after stage b and between the stages corresponding to the images (c) and (d)) is also possible additional pathways of ejection.

Considering all the above parameters, our MD simulations demonstrate the carbon atom transfer from the nanotube to the nickel cluster as a major mechanism for carbon atom ejection (Figure 4). First an electron impact breaks a bond between two three-coordinate carbon atoms near the edge of the hole in the nanotube leading to formation of a carbon chain attached to the nanotube with both ends (Figure 4a). Dissociation of any other bond generates highly unstable carbon atoms that usually reform a new bond within the next MD relaxation stage, while dissociation of the bond between two three-coordinate atoms at the edge of the carbon structure forms a stable chain of carbon atoms, consisting of triple and double bonds, or cumulene double bonds, which formally means that there are no unstable dangling bonds. The nickel cluster appears to stabilize the carbon chains, which is clear from the more than two-fold increase in the average number of carbon chains present in contact with the cluster compared to the number in nickel-free regions (Table 1). The relative stability of two-coordinate carbon atoms in chains is evident by the vast abundance of these atoms compared, for example, to one-coordinate carbon atoms (Table 1). It should be noted that similar carbon chain formation reactions have been frequently observed in irradiation-induced[13,31] or thermally activated[29,54,55] graphene-fullerene transformation processes and for graphene edge etching under the e-beam in the TEM.[56,57] If a chain of two-coordinate carbon atoms is formed and then adsorbed onto the metal cluster, and does not incorporate back to nanotube structure immediately, electron impacts can lead to dissociation of the chain into atoms that are then adsorbed or dissolved into the nickel cluster (Figure 4c). The collision of an incident electron with one of these atoms will then result in its ejection (Figure 4d). Our calculations show that the adsorption energy of carbon adatoms on the nickel surface according to the interatomic potential used[29] is only 6 eV, which is significantly less than the maximum of transferred energy from the 80 keV e-beam to a carbon atom (~16 eV calculated according to the standard theory of elastic electron scattering between a relativistic electron and the nucleus, see eq. (6) of Ref. 13). In general, the reactions of chain formation and chain dissociation should not require a significant energy transfer from the e-beam as they can be achieved through a sequence of steps in which only one carbon-carbon bond is broken at a time. In contrast to the direct ejection of carbon atoms from the nanotube sidewall which requires significant activation energy,[13,30] the pathways enabled by the nickel cluster consist of several successive steps with significantly smaller activation barriers due to the interactions and bonding between carbon and nickel facilitating the entire process.

Additional important mechanisms of carbon atom ejection are based on knock-out of two-coordinate (both in chains and not) and one-coordinate carbon atoms bonded to the metal cluster (Figure 4b, **Table 1**). The contributions of these pathways to the overall ejection rate are minor but not insignificant standing at 12–15% for one-coordinate carbon atoms, 7–10% for two-coordinate carbon atoms in chains and 4–7% for other two-coordinate atoms. Interactions between such carbon atoms and the nickel cluster clearly decrease their ejection threshold energies which is consistent with reported previously *ab initio* calculations demonstrating a decrease in the threshold energies for ejection of carbon atom from a nanotube structure interacting with osmium clusters.[10] This effect was proposed to explain the main mechanism of nanotube cutting by the osmium cluster under electron beam irradiation.[10] Our MD simulations show that in the case of nickel the direct ejection of carbon atom from nanotube bonded to the metal cluster is not the major pathway, as the detachment of individual or pairs of carbon atoms prior the ejection step is more likely to occur, as described above. Thus, it can be deduced that ejection threshold energies for different carbon atoms bonded to the nickel cluster increase (or equivalently ejection cross-sections decrease) in the following order: adatoms, one-coordinate carbon atoms, two-coordinate atoms in chains and other two-coordinate atoms.

The relatively low binding energy (and therefore ejection threshold energy) of carbon adatoms to the nickel cluster leads to a high rate of carbon adatom ejection from the cluster and therefore a low number of adatoms (1.2 – 1.6) are present in the simulations (**Table 1**). Thus the irradiation-induced continuous ejection of carbon adatoms efficiently prevents passivation of the nickel catalyst and contamination of nanotubes with amorphous carbon. In this respect the electron beam plays the analogous role to that proposed of high energy ions in the nanotube growth mechanism in low temperature plasma-enhanced chemical vapor deposition.[21]

The widely accepted mechanism of carbon nanotube growth on nickel clusters was deduced on the basis of atomistic simulations.[15,19,22,23,25,26,28] It involves three key stages: (1) adsorption and diffusion of carbon atoms on the surface of the metal cluster, (2) the formation of polyyne chains, (3) the attachment of the chains to the edge of the carbon nanotube followed by formation of new $sp^2$ structures which lift off from the surface of the cluster. Comparison of the mechanism of carbon atom ejection during the nanotube cutting by the e-beam and the mechanism of nanotube growth reveals a number of interesting similarities. Both mechanisms include virtually identical steps but in reverse order. The surface of the metal cluster serves as a platform for adsorption and ejection from nanotube or adsorption and incorporation in nanotube of carbon adatoms during nanotube cutting or growth, respectively. In both cases the main role of the nickel cluster is to decrease the activation barriers for reactions, such as the dissociation of the $sp^2$ structure into polyyne chains activated by the e-beam in the nanotube cutting process, or the polyyne carbon chain transformation into $sp^2$ carbon atoms activated by heat in the nanotube growth process. The polyyne carbon chains appear to be significant intermediate structures which have been predicted by a number of simulations devoted to high temperature transformation of the open nanotube end[58] and carbon nanotube growth on nickel clusters.[15,19,22,23,25-28]

**Kinetics of nanotube cutting.** As the nanotube structure experiences significant rearrangements during the cutting process, it is interesting to investigate whether the kinetics of the cutting process depends on the size of the system. After a short period of about 500–1000 s or 5–10 ejected atoms from the beginning of the simulations, during which the numbers of different types of atoms and non-hexagonal rings rise to their average values (Figure 2), the average time between the irradiation-induced events does not change considerably (Figure 5a), and can be assumed approximately constant during the whole process. However, fluctuations in the average time between ejection events are much more prominent (Figure 5b).

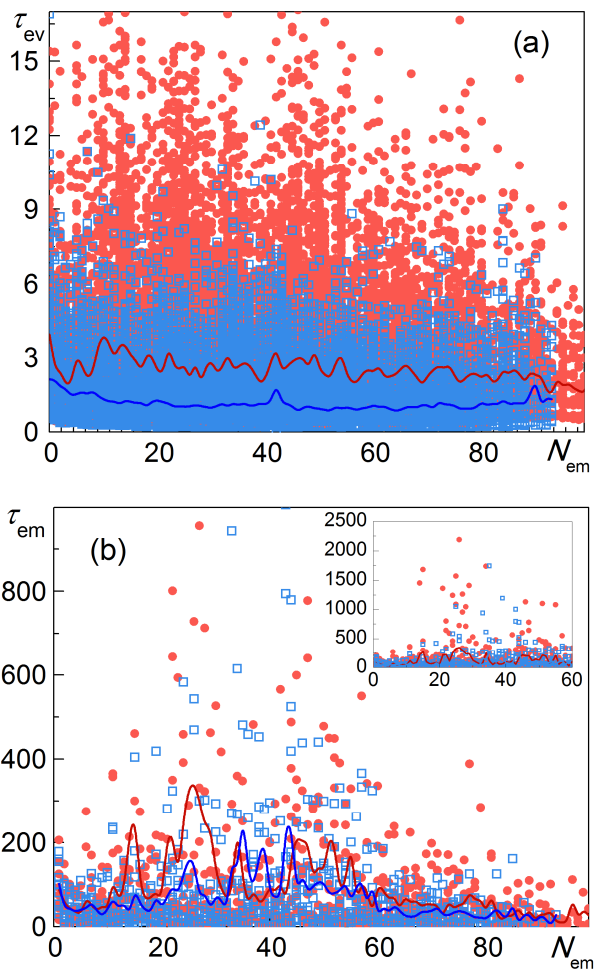

**Figure 5.** Time intervals between (a) all irradiation-induced events, $\tau_{ev}$ (in s), and (b) ejection events, $\tau_{em}$ (in s), obtained in 10 simulation runs of series A (red filled circles) and series B (blue open squares) as functions of the number of emitted carbon atoms, $N_{em}$. Insert: zoomed-

out dependences of time intervals between ejection events, $\tau_{em}$ (in s), on the number of emitted carbon atoms, $N_{em}$, in series A and B. The average values of ejection times in each system size are indicated by the red and blue lines, respectively.

During the initial stages of the simulations, the distribution of the average ejection times is relatively narrow as the structures are all very close to the original structure. These initial average ejection times, based on the ejection times for the first 5 atoms emitted, are estimated to be 50 ± 7 s and 53 ± 7 s in series A and B respectively. After ejection of the first 5 carbon atoms the subsequent ejection times deviate considerably between the different simulations. The scatter of data becomes more prominent with the ejection of more atoms and in some simulations very long ejection times of up to 2000 s are observed in the region of 20-30 emitted atoms. Analysis of the precise structures that manifest such long ejection times reveals that in these structures, the nickel cluster is located on the nanotube sidewall facing towards the electron beam and thus the momentum transferred from impacting electrons to carbon atoms adsorbed or dissolved in the cluster dissipates preferentially along the nanotube sidewall. As a result, in such a configuration, virtually all carbon atoms that are emitted from the nickel cluster are caught by the opposite side of the nanotube and do not leave the system. Though no carbon atoms are ejected in this case, structural rearrangements are still induced in the nanotube and continue until carbon bond network around the nickel cluster is reorganised to a stable configuration. As the nanotube decreases in size, as a consequence of emitted carbon atoms, it is no longer able to screen both the cluster and carbon atoms absorbed or dissolved within the cluster, thus atoms are more readily ejected out of the system over time. Therefore, after the ejection of approximately 70 atoms, nanotube screening effects become negligible and the distribution of ejection times narrows and the average ejection time decreases to 37 ± 3 s and 30 ± 3 s in series A and B, respectively.

To investigate the effect of nanotube diameter on the cutting rate we also performed 10 simulations with parameters identical to series A for a (10,10) nanotube with the same initial hole and adsorbed nickel cluster. Based on the first 5 carbon atoms ejected in these simulations, the initial average ejection time is estimated to be 184 ± 26 s, which is three times greater than for the narrower (5,5) nanotube. Therefore, the curvature of the nanotube sidewall has a strong influence on the cutting rate as the wider nanotube has a stronger interatomic bonding due to lower pyramidalization of the $sp^2$ carbon atoms. We predict that the size of the metal cluster will provide the opposite effect. Bigger nickel clusters should increase the rate of cutting due to increased contact with the nanotube and an increased capacity to adsorb carbon atoms. This may explain the apparent discrepancy in the calculated and experimental cutting rates as the size of nickel nanoparticles in the experiment varies between approximately 50-100 atoms.

**Verification of the CompuTEM algorithm.** The considered process of nanotube cutting is a good example of the type problem which the CompuTEM algorithm[13,30,31] is ideal for, *i.e.* the simulation of structure evolution under electron irradiation in the TEM, and can therefore be used to verify aspects of the algorithm. In particular, it demonstrates the importance of taking into account the relaxation of structures between successive irradiation-induced events. If the duration of the high-temperature MD step that is used to model the relaxation of the structure between irradiation induced events is set at $t_{rel}$ = 0 ps, fast decomposition of the nanotube into chains of carbon atoms is observed (Figure 6). Successive electron collisions lead to wide spread bond breaking and rapid rupture of the nanotube sidewall. However, in reality most of the broken bonds are reformed before the next bond breaking event occurs, induced by irradiation over the timeframe of several seconds, as the bond re-formation reactions have very small energy barriers.[54,55] If a description of the bonds re-formation step is omitted in the simulations, the right axis) as functions of time, $t$ (in s). The moments of time corresponding to structures (a−d) are shown using vertical dashed lines. The minimal transferred energy is $E_{min}^{(2)} = 13$ eV.

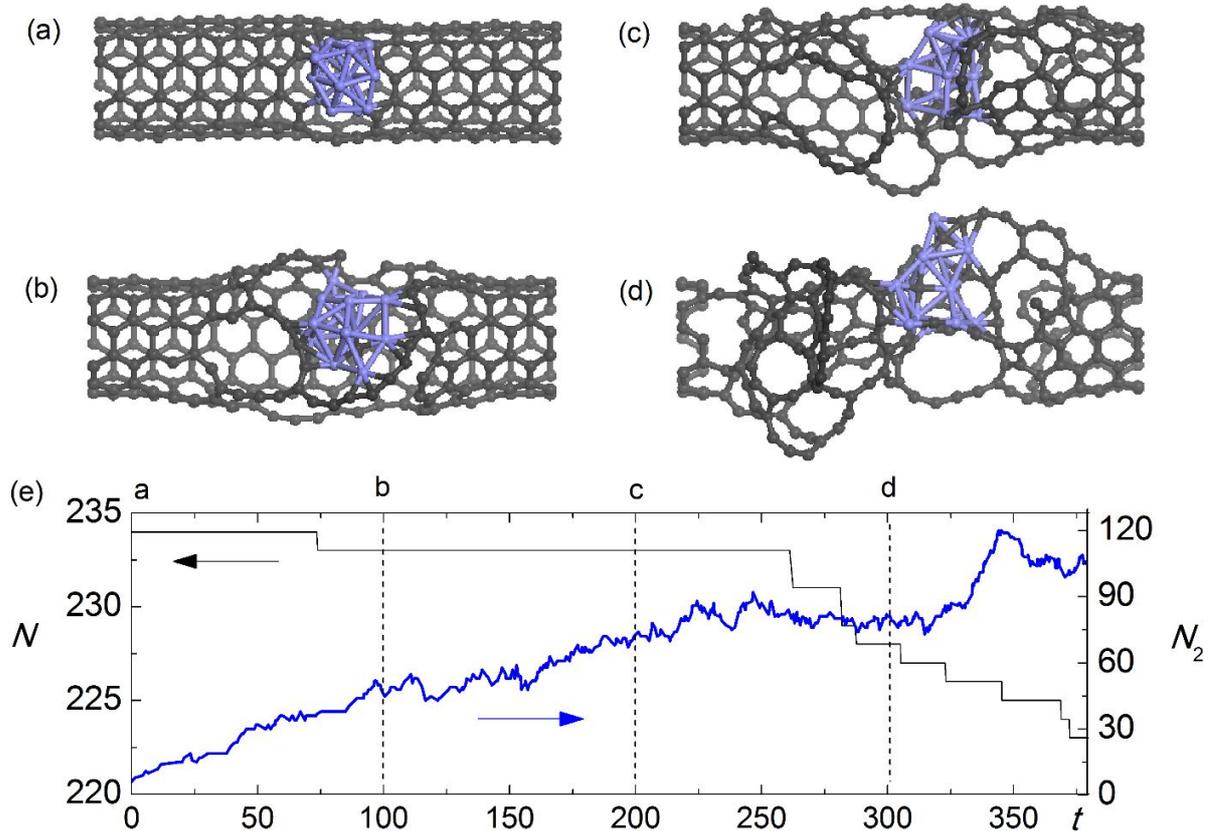

**Figure 6.** (a-d) Evolution of the structure of a carbon nanotube with an adsorbed nickel cluster under irradiation by electrons with a kinetic energy of 80 keV and a flux of 4.1·10$^6$ electrons/(s·nm$^2$) simulated without taking into account structure relaxation between irradiation-induced events: (a) 0 s, (b) 100 s, (c) 200 s, (d) 300 s. The direction of the electron beam is out of the page. (e) Calculated total number of carbon atoms, $N$, in the considered structure (thin black line, left axis) and number of two-coordinate and one-coordinate atoms, $N_2$ (thick blue line,

important reactions that lead to bond reconstruction are excluded completely and the unphysical growth of two and one-coordinate carbon atoms takes place (Figure 6). It should also be noted that a simulation approach which includes relaxation processes within the MD technique is required to obtain a nanotube cap in the simulations of the initial stages of nanotube growth on nickel clusters.[24] The correct choice of the minimal energy transferred to impacted carbon atoms, $E_{min}^{(2)}$, temperature and duration of the relaxation stage, $T_{rel}$ and $t_{rel}$, are also important. For instance, we observed that the temperature of the MD relaxation step of $T_{rel} = 1500$ K (combined with $t_{rel} = 10–30$ ps and $E_{min}^{(2)} = 10–13$ eV) is clearly too low for adequate description of the nanotube cutting process (see Supplementary Information). The results obtained in two series A and B of simulations, with $T_{rel} = 2000$ K, $t_{rel} = 10–30$ ps, and $E_{min}^{(2)} = 10–13$ eV, are consistent with the experimental observations and with each other, giving qualitatively the same structures and very similar quantitative results. The quantitative difference in the kinetic characteristics of the cutting process in these two series does not exceed 50%, confirming that the parameters used in these two series of simulations are adequate for the description of structure relaxation between successive irradiation-induced events (see details in Supplementary Information). Most importantly, the results of the theoretical CompuTEM approach correlate well with the experimental AC-HRTEM measurements (Figure 1), correctly predicting the key stages in the nanotube cutting process by the e-beam facilitated by nickel clusters.

CONCLUSION

Clusters of nickel atoms are shown to catalyze the cutting of single-walled carbon nanotubes initiated by 80 keV electrons of a TEM. Real time image sequences show the loss of carbon atoms from the nanotube sidewall and the formation of large defects which is followed by the reorganization of the carbon framework to form two end caps on the two segments of nanotube finally resulting in a complete severing of the nanotube. The nickel clusters are observed to play a crucial role in the process stabilizing the defects and edges of the nanotube and catalyzing the formation of the closed fullerene caps. This method enables to create a gap between the ends of the cut nanotubes in a reproducible manner without chemical contamination, with the size of the gap controlled by the size of the nickel cluster of ~1 nm, which is important for the fabrication of nanodevices.

Molecular dynamics simulations based on the CompuTEM algorithm[13,30,31] confirm cutting of the (5,5) nanotube by a nickel cluster adsorbed on the hole in the nanotube sidewall prior to the cutting process. Detailed analysis of the mechanism reveals that the cutting by the 80 keV electron beam takes place over a timeframe of approximately $10^4$ s and involves the ejection of 100 atoms. The cutting is shown to proceed *via* the four key stages stages: (1) narrowing of the nanotube, (2) unfolding of the narrow part of the nanotube into a graphene nanoribbon, (3) the formation of chains of carbon atoms connecting two fully closed nanotube ends and finally (4) the complete separation of the nanotube segments.

Both the assistance of the nickel cluster and the electron beam are demonstrated to be crucial for nanotube cutting to occur. In the absence of the nickel cluster, the rate of carbon atom ejection under electron irradiation is found to decrease by an order of magnitude. In addition, the nanotube without metal undergoes completely different structural transformations under the e-beam, including the formation of metastable structures with reconstructed vacancy defects (with very few or no two-coordinate atoms) in the nanotube sidewall. Moreover, simulations for the nanotube with a nickel cluster but no electron beam demonstrated no carbon ejection or even carbon atom dissolution into the nickel cluster. Thus, both electron irradiation and the presence of a metal cluster are crucial factors in the cutting of nanotubes.

Detailed analysis of local structural changes showed that about 80% of events induced by the electron beam (dissociation and rearrangement of chemical bonds, ejection of carbon atoms etc.) take place involving carbon atoms not strongly bonded to the nickel cluster but located in the non-perfect regions of nanotube structure in the vicinity of the metal. The key reactions involve the breaking of bonds of three-coordinate atoms in non-hexagonal rings, initiating reconstruction of defective parts of the nanotube and the formation of nanotube caps. Furthermore, about 64% of *ejected* carbon atoms are adatoms adsorbed on the nickel cluster, indicating that the main pathway of nanotube cutting by the electron beam involves total dissociation of bonding of a carbon atom with nanotube followed by its adsorption on the metal particle. Ejection of two-coordinate and one-coordinate carbon atoms bonded to the nickel cluster also occurs but less likely. The dominant mechanism of carbon atom ejection thus can be described as follows: (1) the formation of a chain of two-coordinate carbon atoms attached by both ends to the carbon nanotube and simultaneously bonded to the nickel cluster, (2) dissociation of the chain into atoms adsorbed or dissolved in the cluster, (3) the knocking out of carbon adatoms from the nickel cluster. Accounting for the additional knocking out of two and one-coordinate carbon atoms at all stages of this mechanism would enable a complete description of nanotube cutting processes.

The rate of carbon atom ejection is found to vary strongly with time, with the time interval between ejection events of about 50 s at early stages of the cutting process which potentially can increase by orders of magnitude during the intermediate stages if the position of nickel cluster allows recapturing the ejected carbon atoms by the nanotube. The carbon atom ejection rate accelerates again towards the end of the cutting process, when only chains of carbon atoms remain bridging between two severed segments of the nanotube. A two-fold increase in the nanotube diameter has a drastic effect on the kinetics of cutting, reducing the rate of carbon atom ejection by a factor of three.

In a wider context this study reveals very important information about the role of the metal catalyst in nanotube cutting. Previous experimental evidence shows that Os is a better catalyst for nanotube cutting compared to W and Re from the same period VI.[10] We also observe that Os is a better catalyst for nanotube cutting in comparison with Fe and Ru from the same group VIII. We believe that this is related to the balance between the strength of metal-carbon σ-bonds and the cohesive energy of the individual metal clusters. The simulations performed in this study for nickel clusters confirm this conclusion. Firstly, strong bonding between metal clusters and the edge or defects of the sp$^2$ carbon structure weakens the nearby carbon-carbon bonds and thus promotes irradiation-induced reactions and rearrangement of the carbon structure and also carbon atom ejection. Secondly, the high cohesive energy of the metal cluster correlates with weak bonding between the metal cluster and any single carbon atoms or polyyne chains. Such weak bonding promotes the irradiation-induced cleaning of the catalyst *via* knocking out of the weakly bound carbon from the metal cluster. As a number pathways in which carbon atoms are ejected from the nickel catalyst under the electron beam are revealed at the atomic level in the performed simulations, it is possible that etching of carbon nanostructures catalyzed by different transition metals could have different atomistic mechanisms, which must be explored in the future by a combination of AC-HRTEM observations and CompuTEM simulations.

Our methodology and the atomic level understanding of metal cluster and nanotube interactions under electron beam irradiation paves the way towards the precise control and manipulation of carbon nanostructures which can unlock their full potential for practical applications at the nanoscale. In particular, one may think of such a technologically relevant process as the longitudinal cutting of carbon nanotubes and graphene. In the first stages of our simulations when the nanotube still retains its integrity (before formation of graphene ribbons and carbon chains) no preference is observed for longitudinal or circumference cutting, i.e. the cluster moves in an arbitrary direction on the nanotube sidewall. However, we suggest that controlling the position of the transition metal cluster, for example, with a STM or AFM tip, will make it possible to guide the trajectory of the cluster and enabling cutting of carbon nanotubes or graphene in a pre-determined fashion. This can be a step forward in the controlled design of graphene flakes and nanoribbons.

ASSOCIATED CONTENT

**Supporting Information**. Movies based on experimental observations of nickel catalyzed electron beam assisted cutting of carbon nanotubes in AC-HRTEM and CompuTEM simulation of nickel catalyzed electron beam assisted cutting of the (5,5) nanotube (from series B of simulations), detailed results on e-beam irradiation of pristine nanotubes and thermal treatment of carbon nanotubes with nickel clusters, simulation results for a MD relaxation step temperature of 1500 K and discussion of sensitivity of the results to simulation parameters.

AUTHOR INFORMATION

**Corresponding Author**

*Address correspondence to liv_ira@hotmail.com.

ACKNOWLEDGMENT

AP and AK acknowledge the Russian Foundation of Basic Research (14-02-00739-a). IL acknowledges support by the Marie Curie International Incoming Fellowship within the 7th European Community Framework Programme (Grant Agreement PIIF-GA-2012-326435 RespSpatDisp), Grupos Consolidados del Gobierno Vasco (IT-578-13) and computational time on the Supercomputing Center of Lomonosov Moscow State University[59] and the Multipurpose Computing Complex NRC "Kurchatov Institute."[60] TWC and ANK are supported by the European Research Council (ERC) and the Engineering & Physical Sciences Research Council (EPSRC).

# Supplementary Information
## Atomistic mechanism of carbon nanotube cutting catalyzed by nickel under the electron beam

*Irina V. Lebedeva, Thomas W. Chamberlain, Andrey M. Popov, Andrey A. Knizhnik,
Thilo Zoberbier, Johannes Biskupek, Ute Kaiser and Andrei N. Khlobystov*

S1. EDX spectra for bundles of filled nanotubes

EDX spectra were recorded for small bundles of SWNTs (3-10 nanotubes) filled with each metal on a JEOL 2100F TEM equipped with an Oxford Instruments X-rays detector at 100 kV.

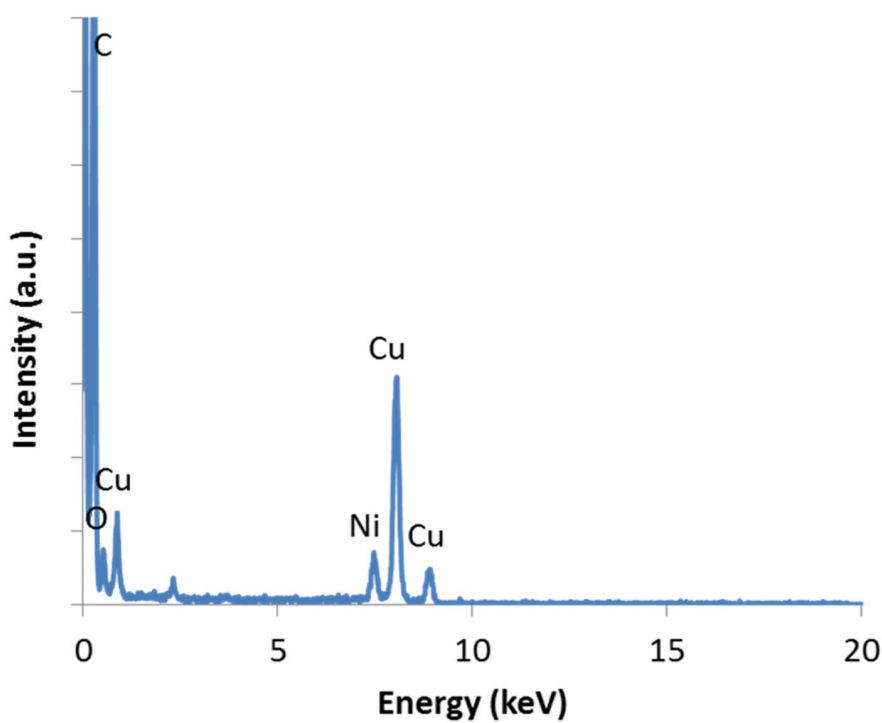

**Figure 1S.** The EDX spectrum of Ni-NPs@SWNT bundle (containing approximately 80 nanotubes) confirming the presence of nickel within the nanotubes. The Cu EDX peaks are due to the copper TEM specimen grid.

S2. Modeling of e-beam irradiation of pristine nanotubes and thermal treatment of carbon nanotubes with nickel clusters

To verify that the combination of both the metal cluster and electron irradiation is necessary to initiate nanotube cutting we performed supplementary simulations in the absence of each of these factors individually. Simulations for the pristine nanotube of 28 Å length with the same initial hole in the nanotube wall but without an absorbed metal cluster (Figure 3a) showed that carbon atom ejection is much rarer. The average time interval between ejection events is 1600 ± 400 s in 5 simulations using the same parameters as in series A and exceeds 1100 s in 5 simulations with the same parameters as in series B. This ejection rate is more than twice as slow as the rate observed in the simulations of irradiation-induced graphene-fullerene transformation[S61] and is an order of magnitude smaller than the ejection rate in the presence of the nickel cluster. A total of 16 carbon atoms were emitted in all simulations of the pristine nanotube in a duration of 2000 – 5000 s. Analysis shows that 7 of the emitted atoms were two-coordinate atoms in non-chain structures, 4 were one-coordinate atoms, while the rest consisted of a mixture of three-coordinate atoms in non-hexagonal rings, three-coordinate atoms neighbouring with two and one-coordinate atoms or non-hexagonal rings, and two-coordinate atoms in chains. The evolution of the structure of the pristine nanotube under electron irradiation is shown in Figure 3.

The following changes are induced in the pristine nanotube upon electron irradiation. First formation of non-hexagonal rings around the hole takes place (Figure 3b, Figure 3g). This can lead to curing of the hole in the nanotube wall (Figure 3c) that is accompanied by local deformation of the nanotube and manifests itself as a decrease in the total number of one and two-coordinate carbon atoms (Figure 3g). However, such states are not very stable and within the order of 100 s the nanotube is reconstructed (Figure 3d-f). The number of non-hexagonal rings increases again and these rings spread along the nanotube axis away from the hole (Figure 3d-g). The number of one and two-coordinate carbon atoms also increases and exhibits significant fluctuations (Figure 3g), demonstrating opening and closing of holes in the nanotube wall (see a hole on the reverse side of the nanotube in Figure 3f). In addition to non-hexagonal rings, bridges between opposite sides of the nanotube are formed by individual or chains of two-coordinate carbon atoms (Figure 3e and f). Therefore, simulations of the irradiation of pristine nanotube demonstrate that in the absence of the nickel cluster structural reconstruction induced by the e-beam proceeds via alternative pathways. In particular, the ejection rate of carbon atoms decreases drastically in the absence of the nickel cluster.

Simulations of the same initial structure with a nickel cluster adsorbed on the hole in the nanotube wall at temperature 2000 K but without electron irradiation did not reveal any structural rearrangements within tens of nanoseconds. In particular, no atom ejection is observed in such simulations. The duration of these simulations corresponds to the combined high-temperature relaxation steps of simulations with electron irradiation after 300 – 1000 "successful" electron collisions. Therefore, it is clear that the introduction of the high-temperature steps in order to allow relaxation of the structure between successive structure changing electron collisions does not directly affect the ejection rate. In addition, these simulations along with the simulations of electron irradiation of the pristine nanotube demonstrate that a combination of both electron irradiation and a metal cluster, are required to initiate nanotube cutting using low kinetic energy electrons (80 keV).

## S3. Simulation results for relaxation temperature 1500 K

To investigate the effect of relaxation temperature $T_{rel}$ we have performed series of simulations A2 and B2 with the parameters $t_{rel}$ and $E_{min}^{(2)}$ similar to those of series A and B but with the temperature of the relaxation step of $T_{rel} = 1500$ K (Table 1S). Series A2 and B2 contain 5 simulation runs each. These simulations are interrupted before the nanotube cutting. Nevertheless, their duration is sufficient to make comparison with the results presented above for temperature $T_{rel} = 2000$ K.

In these simulations, the system is shown to pass through the same intermediate stages and the mechanism of nanotube cutting is qualitatively the same as in series A and B. The average times between irradiation-induced events and ejection events do not significantly alter as a product of changing the temperature (Table 1S). However, the relative abundance of two-coordinate carbon atoms in chains and one-coordinate carbon atoms is increased drastically compared to the results at temperature $T_{rel} = 2000$ K, especially in the simulation runs with short relaxation times, $t_{rel} = 10$ s. Therefore using a lower temperature during the relaxation MD stage of $T_{rel} = 1500$ K does not adequately describe irradiation events caused by electrons with the kinetic energy of 80 keV with the flux of $10^6$-$10^7$ e$^-$/nm$^2$/s at room temperature. In addition to testing the simulation parameters, these results qualitatively describe experiments with either increased electron flux or higher temperature, revealing that electron flux and sample temperature can potentially be used to the level of imperfections with structures.

**Table 1S.** Calculated average time between different irradiation-induced events, their relative frequencies for impacted carbon atoms of different types and the number of atoms of different type averaged over total time for the simulation of nickel catalyzed electron beam assisted nanotube cutting with a MD relaxation step temperature of $T_{rel} = 1500$ K.

| Irradiation-induced events | All | | Ejection | | Number of atoms[a] | |
|---|---|---|---|---|---|---|
| Simulation series | A2 | B2 | A2 | B2 | A2 | B2 |
| Minimal transferred energy $E_{min}$ (eV) | 13 | 10 | 13 | 10 | 13 | 10 |
| Relaxation time $t_{rel}$ (ps) | 30 | 10 | 30 | 10 | 30 | 10 |

| Total number of irradiation-induced events | 12248 | 11792 | 336 | 150 | | |
|---|---|---|---|---|---|---|
| Average time between ejection events $\tau$ (s) | 2.59 ± 0.05 | 0.99 ± 0.03 | 94 ± 7 | 74 ± 9 | | |
| Atom types | Not bonded to the cluster | | | | | |
| One-coordinate atoms | 0.0032 | 0.0014 | 0.0387 | 0.0200 | 0.07 | 0.07 |
| Two-coordinate atoms except atoms in chains | 0.0262 | 0.0304 | 0.0387 | 0.0267 | 4.98 | 7.98 |
| Two-coordinate atoms in chains[b] | 0.0163 | 0.0174 | 0.0238 | 0.0067 | 3.12 | 4.05 |
| Three-coordinate atoms in non-hexagonal rings | 0.5296 | 0.557 | 0.0268 | 0.0067 | 82.85 | 91.36 |
| Three-coordinate carbon atoms in hexagons | 0.1538 | 0.1771 | 0.0030 | 0.0067 | 170.37 | 172.49 |
| Total for atoms not bonded to the cluster | 0.7291 | 0.7830 | 0.1310 | 0.0667 | 261.39 | 275.96 |
| | Bonded to the cluster | | | | | |
| One-coordinate atoms | 0.0287 | 0.0355 | 0.1994 | 0.2600 | 1.04 | 1.59 |
| Two-coordinate atoms except atoms in chains | 0.0501 | 0.0381 | 0.0417 | 0.0400 | 4.52 | 4.40 |
| Two-coordinate atoms in chains[b] | 0.0492 | 0.0923 | 0.1369 | 0.2133 | 5.99 | 9.00 |
| Three-coordinate atoms in non-hexagonal rings | 0.0079 | 0.0119 | 0.0030 | 0 | 0.55 | 0.64 |
| Three-coordinate carbon atoms in hexagons | 0.0026 | 0.0044 | 0 | 0 | 0.17 | 0.32 |
| Adatoms | 0.0682 | 0.0325 | 0.4673 | 0.4133 | 1.14 | 1.12 |
| Ad-dimers | 0.0038 | 0.0024 | 0.0208 | 0.0067 | 0.05 | 0.08 |
| Total for atoms bonded to the cluster | 0.2105 | 0.2170 | 0.8690 | 0.9333 | 13.46 | 17.15 |

[a] Fixed atoms at the nanotube edges are not counted.
[b] Two-coordinate carbon atoms which have at least one bond with two-coordinate or one-coordinate carbon atoms.

## S4. Sensitivity of results to simulation parameters

In the following we address the choice of parameters for description of structure relaxation between successive irradiation-induced events. Since it is not possible to perform direct atomistic simulations of structure evolution between irradiation-induced events that take seconds under real conditions, a common approach is to accelerate its kinetics by increasing temperature. However, care should be taken in selecting adequate parameters for this simulation step. First of all, to separate the effects of electron irradiation and high temperature the temperature $T_{rel}$ and duration $t_{rel}$ of this high-temperature stage should be chosen so that thermally induced transformations of the pristine structure are virtually excluded, i.e. the following condition should be fulfilled

$$N_{ev} t_{rel} \ll t_{th},$$

where $N_{ev}$ is the number of irradiation-induced events during the simulation of the irradiation-induced process and $t_{th}$ is the characteristic simulation time required for the thermally induced process analogous to the considered irradiation-induced process to take place at the elevated temperature $T_{rel}$. The simulations at high temperature without electron irradiation have shown no indication of nanotube cutting and even atom ejection. Therefore, this condition is clearly fulfilled for the considered process and the simulation parameters $E_{min} = 10-13$ eV, $t_{rel} = 10-30$ ps and $T_{rel} = 1500-2000$ K.

The restriction on $t_{rel}$ and $T_{rel}$ from the other side is that they should be sufficiently large to describe recovery of bonds broken in the result of electron collisions. The results obtained at relaxation temperature $T_{rel} = 1500$ K show that this temperature is insufficient to describe structure relaxation between successive irradiation-induced events even at duration of the relaxation step of $t_{rel} = 30$ s. The relative abundance of two-coordinate carbon atoms in chains and one-coordinate carbon atoms is increased drastically as compared to the results at relaxation temperature $T_{rel} = 2000$ K, especially in series B2 with the short relaxation time $t_{rel} = 10$ s (both for the carbon atoms in the contact with the nickel cluster and not, Table 1S). Consequently, the contribution of these types of atoms to the irradiation-induced events increases (see Table 1S).

The choice of higher relaxation temperature $T_{rel} = 2000$ K does not eliminate all restrictions on $t_{rel}$. In particular, the reactions reverse to the formation of carbon chains at the edge of the carbon network (similar to the process shown in Figure 4a and b), though they have a very low barrier, also have a large characteristic time in the pre-exponential factor that can be on the order of 10 ps.[S62,S63] Therefore, $t_{rel} \geq 10$ ps should be chosen to prevent unphysical generation of carbon chains. In addition to increasing the relaxation time $t_{rel}$, the excessive generation of chains can be suppressed by increasing the minimal transferred energy $E_{min}$ for the corresponding types of atoms. Thus, it can be expected that structure relaxation between irradiation-induced events is somewhat better treated in series A of simulations. Indeed the total number of two-coordinate carbon atoms in chains averaged over time is greater by 20% in series B than in series A.

Nevertheless, the results are qualitatively similar in series of simulations A and B and are even quantitatively close in these two series. The relative frequencies of all irradiation-induced events and atom ejection events for different types of atoms are almost the same in the both series of simulations (Table 1). The number of carbon atoms in chains and, correspondingly, the number of carbon adatoms averaged over time are greater in series B than in series A only by 30%. The time between atom ejection events is, correspondingly, 30% greater for series A than for series B. The time of complete cutting of the nanotubes in the finished simulations is one a half times greater in series B. This discrepancy in quantitative results of series A and B is acceptable and demonstrates that the simulation parameters approach the full description of the effect of electron irradiation.